# Innovative Low Cost Laboratory Automation Environment and LabVIEW Reformation Application Case Study

Garnet Cameron, David Shiner
University of North Texas

## Abstract
In recognition of the catalytic role of instruments, we report on an original, low-cost, robust, LabVIEW-based automation development environment configuration and application to reformation of a legacy laser atomic spectroscopy system. Open source, version and configuration control, full back-up, and remote/distributed capability characteristics make the new environment 500% better. System reformation using reusable type definitions, functional encapsulation, increased modularization, and polymorphism boosted performance 983%. Both the environment configuration and reformation strategies are transferrable to most endeavors.

## 1 Background

Experimental work focuses on the relevant central phenomenon or measurement. However, instrumentation and development infrastructure are key components that facilitate the end goal. A significant constraint is often the budget. These factors motivate an innovative approach to both reduce cost while providing state of the art capability within the context of a legacy system. Several shareware and minimal commercial components are integrated with pivotal tuning to achieve useful performance.

LabVIEW automation control via the main Virtual Instrument (VI) called LineShape is used to generate overnight run data which is stored in output files, as depicted in Figure 1. Mathematica scripts analyze statistically the LineShape output data files.

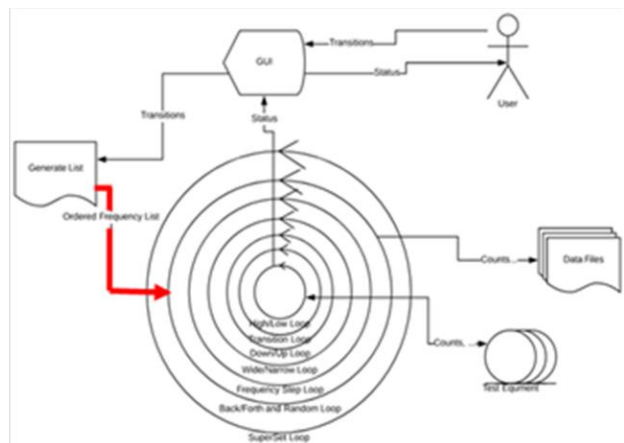

Figure 1 Legacy LineShape High Level Schematic

### 1.1 Legacy Development Environment

The legacy work environment i.e. LabVIEW and Mathematica file version management, file backup, Operating System (OS) support, and equipment of this experiment facilitated a single experimenter. Version control was renaming files, backup was not present[1], only

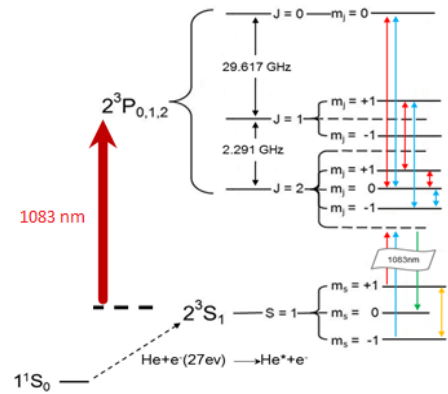

Figure 2 $^4$He Energy Levels

---

[1] RAS file copying to another disk existed previously, but is no longer active.

Windows 7 support existed, and tools and equipment were not centrally organized or scheduled. Introducing a team of experimenters to such an environment can only spell chaos. This cycle supports three PhD candidates and several undergraduates.

## 1.2 Experiment

Our current major work intends to increase the precision of Helium-4 (and Helium-3), n = 2, triplet transition measures as exampled in Figure 2, by at least an order of magnitude over current values [1] [2] [3] [4] via statistical error reduction (increasing N). Improved transition precision tests Quantum Electrodynamics (QED) theory to a higher degree or order, which may foster theoretical enhancement of the Standard Model.

Increased-precision modifications to this experiment lay the foundation for increased precision in associated measures including:

- Isotope shifts of $^3$He versus $^4$He; and
- Determination of:
    - $1/\alpha$; and
    - Nuclear size (Isotope-Shift Method)

Despite applying many feedback stabilization techniques and systems to all known exacerbating variables, slow drifts and small oscillations still occur during measurements. Statistical analysis of $10^4$ iterations of each transition measure allows identification and minimization of leading exacerbating variables permitting approach to the Poisson error limit - $\sigma/\sqrt{N}$ [5]

# 2 Improved Environment

A discussion of the various environment aspects follows in importance order with Backup and Version Control at almost the same significance. Cost weighted highest among the deciding factors in all dimensions.

## 2.1 Backup

Without the safety net of a backup strategy, catastrophic hardware loss could unwind this work many years or decades. Of the many options available in this domain ranging from scheduled tape or disk copy to cloud storage, cost and stability favored the latter. Many cloud storage providers make GBs available free of charge (introductory offer). The experiment softcopy footprint (data, processing tools, and experiment control) is only 3 GB at this point – less than the free quotas available. Google Drive/Backup and Synchronize was selected with 5 GB of free and reliable (hot synchronized and UPS) storage.

The version control approach infers the data storage format.

## 2.2 Version Control

National Instruments LabVIEW allows flexible experiment control via a VI concept. VIs are graphically encoded in proprietary format and stored on disk. Mathematica notebooks are ASCII formatted. Historical/version tracking of the VI and notebook files is important for project stability allowing option management as different approaches are tried.

General version control[2] is available from many products. Our Windows-based Experiment PC constrains the still numerous options [6]. However, a low or no cost requirement presents only a few that boast stable history and are familiar i.e. GNU SubVersion (SVN) and GIT. SVN has many clients/Graphical User Interfaces (GUIs) making it very accessible to users of any skill level.

However, a race condition occurs between SVN temporary file generation upon Commit and Google Drive's dynamic synchronization. In fact, large Commits cause repository corruption. This can be avoided by synchronizing during development idle times – in our case overnight was selected. Google Drive automatic invocation at machine start is disabled. A scheduled task that invokes the Google Drive process at the synchronization window start is paired with another the stops it at the end of the window[3].

## 2.3  New Environment

Figure 3 portrays the resulting architecture. Many improvements are possible including creating a SVN Server. Note that data files are `directly' housed in Google Drive i.e. not versioned. A symbolic link directory complies with LineShape developed to use a local DATA directory.

The salient features of the environment configuration are:

- Google Drive (GD) backup (5 GB free[4])
  As common with the many cloud data services today, code and data catastrophe survivability is ensured. Reinstallation of the GD and SVN repository followed by project (/Experiment) SVN check-out recreates a complete system at the last commit version.

- SVN version and configuration control (freeware available)
  State of the art version configuration control provides durable development like version rollback, version comparison, etc. National Instruments (NI) LV Compare tool is need for proprietary encoded .vi files differencing.

- GD distribution
  Shared workflow is possible by GD installation on multiple machines. In LineShape's case, Data is available for distribution.

- SVN distribution
  Team development is enabled by the SVN Server. Experiment Control (LabVIEW) and Processing (Mathematica) codes become available on multiple machines with modern conflict protections via user accounts and check-in processes.

---

[2] Lab View publishes support for expensive third party proprietary version control systems e.g. Perforce, MS SourceSafe, and Rational ClearCASE (to name a few).
[3] See https://www.wintips.org/how-to-schedule-google-backup-and-sync/
[4] This quantity is expected to grow over time.

- Dynamic Data update
  GD's dynamic synchronization scheme makes data generated in on-going experiments available across distributed machines which is a useful tool in tracking experiment progress.

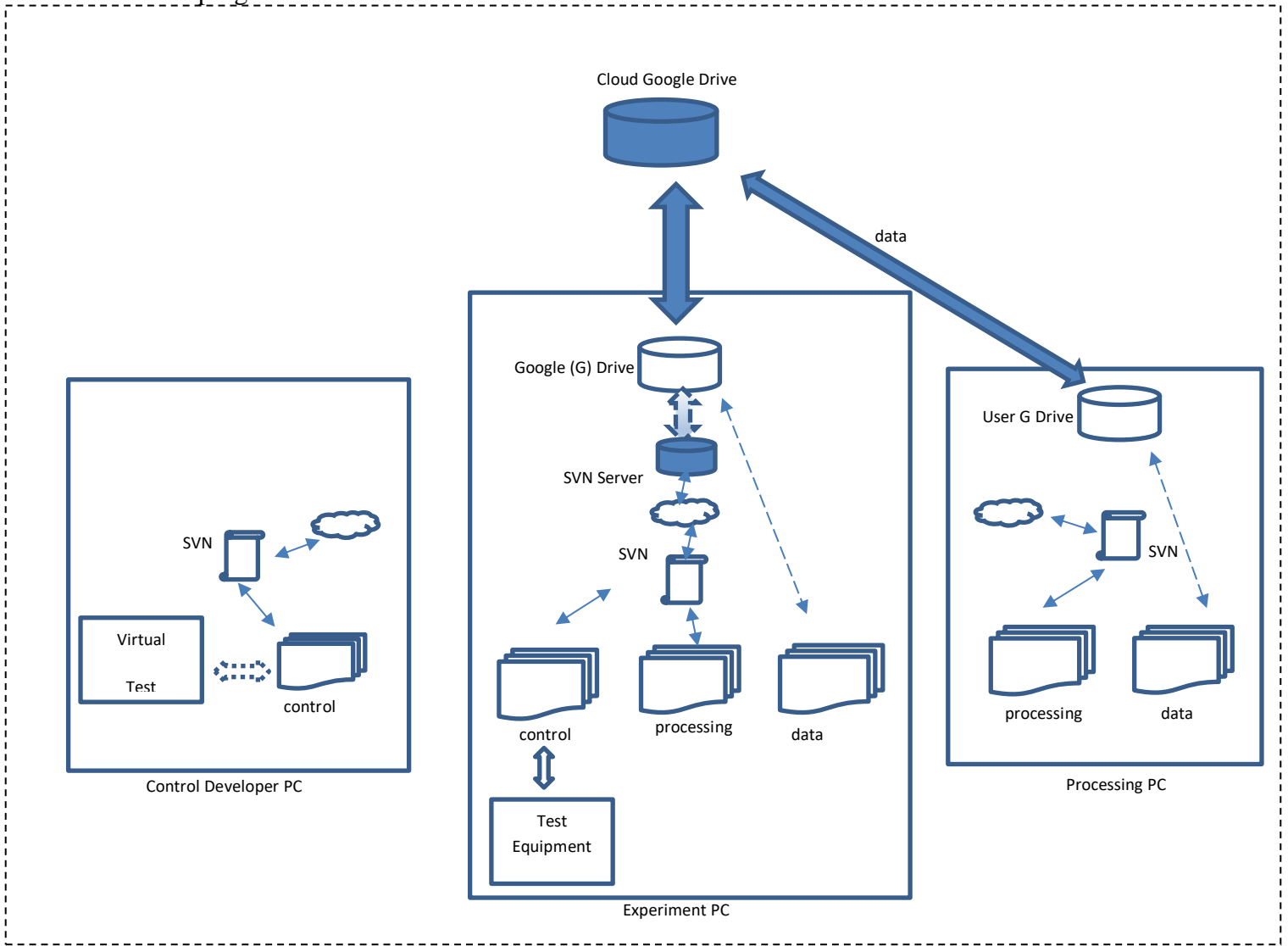

**Figure 3 New Environment Schematic**

## 3 Improved LineShape

All projects begin with proof-of-concept and prototyping. Long-lived endeavors must mature beyond those stages which requires revisiting the fundamental structures and organization that was implemented during prototyping. Often, re-organization and rebuilding is mandated to attain goals of maintainability, flexibility, and simplicity. Failure to mature any system results in arrangements that are awkward, difficult to maintain, and modify which only get worse over time. The high-level architecture of LineShape is shown in Figure 1.

Although LabVIEW-based LineShape has been used and modified for more that fifteen years, the maturation process has not been applied resulting in several architectural and design issues. An extensive, spaghetti-code implementation results as seen in Figure 4.

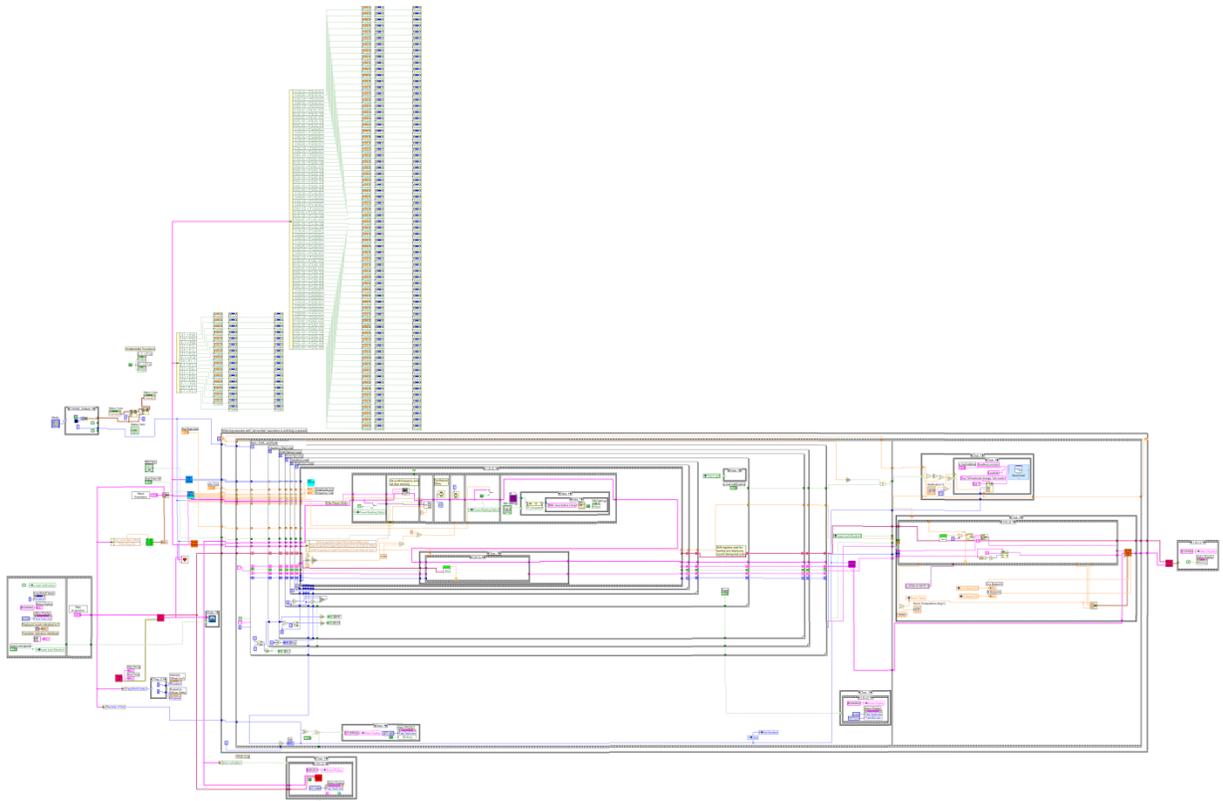

Figure 4 LineShape Extensive and Spaghetti Block Diagram ("zoomed" out)

Significant implemented improvements are discussed below [7].

## 3.1 Reusable Type Definitions

LabVIEW Type definitions[5] create easily reusable structures that are positioned to reduce errors, speed modifications[6], and increase readability (via Unbundle By Name). Any heavily reused structure/cluster should be type-defined. Unfortunately, heavy use of copy and paste has been used in LineShape for cluster propagation as modelled in Figure 5 (a) and (b). The latter shows fragility as breakage occurs with Bundle extension/modification due to structure mismatch. Significant examples are:
- File & Equipment cluster
  This cluster is heavily used carrying identifiers for all test equipment and files used in LineShape.

---
[5] Similar to #include typedef in C or C++.
[6] Via Apply Changes that triggers propagation and recompile in LabVIEW.

- Input Select Frequency cluster
  All user input is hosted in this cluster making it a heavily used grouping. It includes the transition selected, frequency settings, and other user input parameters.

To repair this situation, a typeDef is created for each high-use cluster and swapped into the existing .vi's as shown in Figure 5 (c) and (d). Fragility upon extension does not exist.

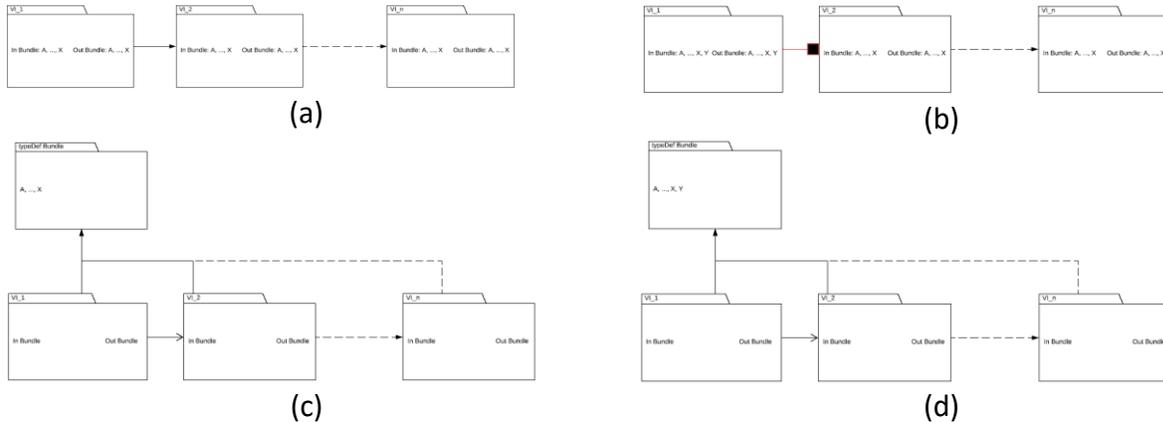

Figure 5 TypeDef Solution: (a) Copy Approach; (b) Breakage Upon Modification; (c) Common TypeDef; (d) Seamless Extension

. New development creates custom controls as exampled by:
- Resolution ring (Microwave Counter)
  Microwave counter resolution is defined as a ring enumeration. Microwave counting/triggering, reading, and comparison is implemented in distributed fashion requiring reuse of the Resolution control.
- ModeType (GPIB Totalizer)
  A GPIB Totalizer/counter can operate in TimeOut or ServiceReQuest modes. The former allows the counter to address/handle zero counts. A radio button implementation is used. Triggering and reading distributed implementation forces reuse.

## 3.2 Functional Encapsulation

A critical characteristic of compact and resilient system architecture is localization of functionality within isolated units. Reuse, maintenance, and stability is enhanced by this approach. LineShape displays atomic function splitting and multiple function combination issues ranging from core functions e.g. generation and iteration of the frequency list, to utility functions e.g. setting microwave frequency and probe laser power.

Encapsulation has been improved with:
- Frequency and Power Separation
  Counts and transition frequency center are dependent on the probe intensity/power. In particular, a systematic power shift is extrapolated to zero power by probing at high and low power levels. Historically changing frequency and power was implemented in the same VI. Splitting these separate functions allows better timing control as shown in Figure 8 (vs Figure 6).

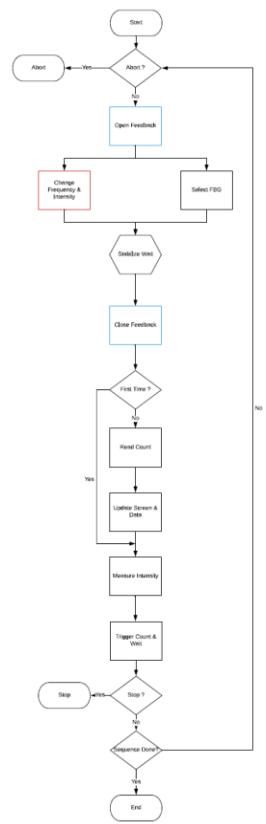

Figure 6 Original Base Loop Flow Chart

- Frequency List Generation and Iteration
LineShape's overall architecture rests on generation of an *apriori*, ordered superset list of frequencies that contains all permutations that will be repeated. This list is generated by a looping implementation without power step, frequency High/Low, frequency Wide/Narrow, High First/Narrow First, or High First/Low First status information for each frequency point as shown in Figure 7. Nesting of the list allows rediscovery of this information during iteration, but results in a duplication of the loops used during creation as seen Figure 4.

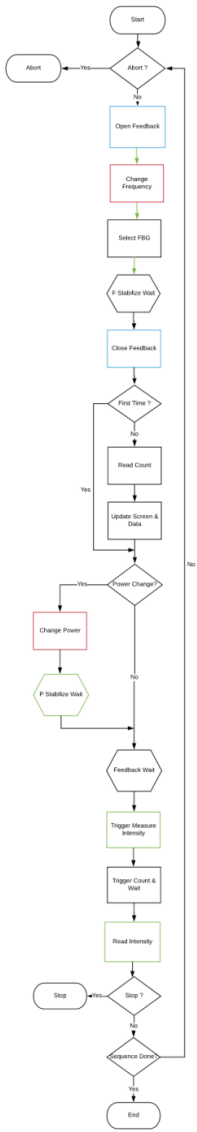

**Figure 8 Frequency & Intensity Separation**

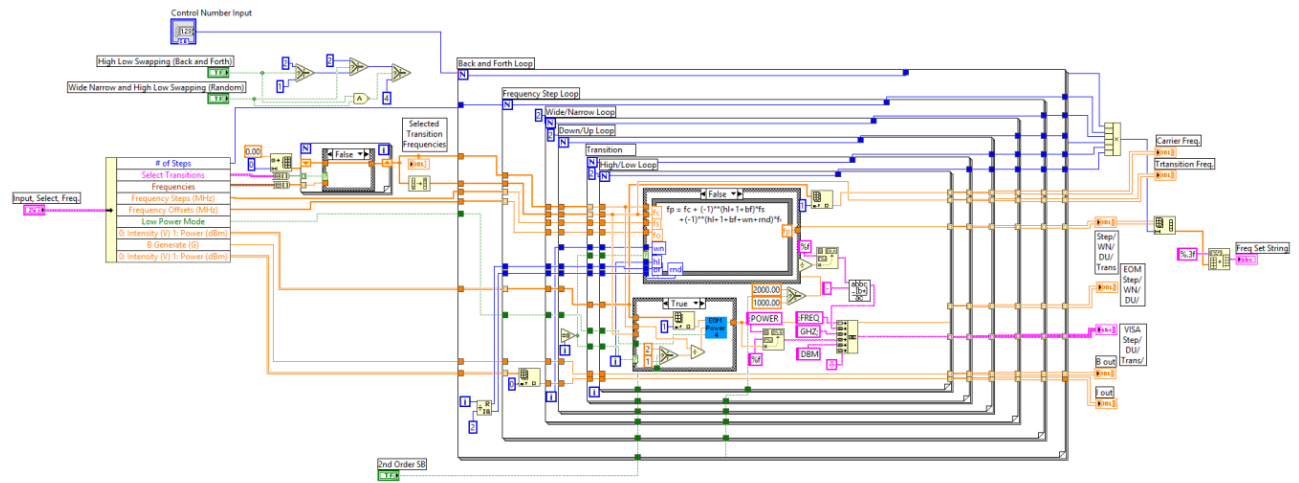

**Figure 7 Nested Frequency List Generation**

One can argue it is just an encoding scheme. However, it exposes the generation implementation to the iteration implementation (experiment run) and more importantly complicates the iteration unnecessarily – a violation of functionality encapsulation. Therefore, any changes in list generation will necessitate changes in iteration.

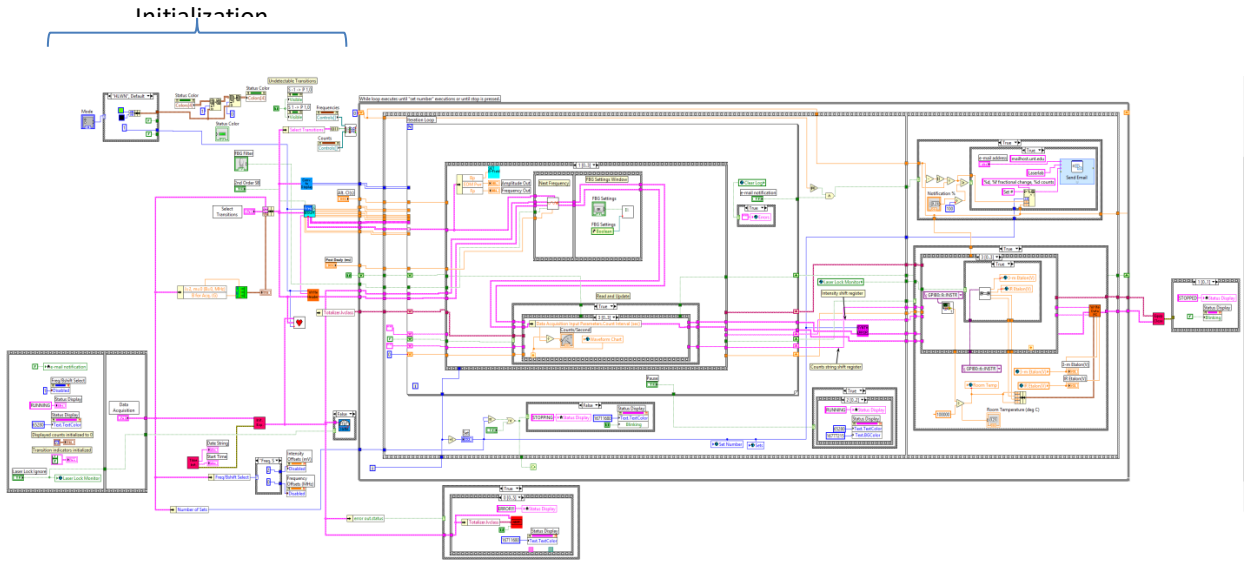

Figure 9 Current LineShape (initialization stage indicated)

A more isolating/functional encapsulated approach would combine the status and frequency information into the generated list producing a "flat" and order-neutral list. Iteration only moves from frequency to frequency presenting blindly status information that is already captured during generation. Simplification results in code of Figure 9 and high level diagram in Figure 10, where only the iteration and superset loops are necessary.

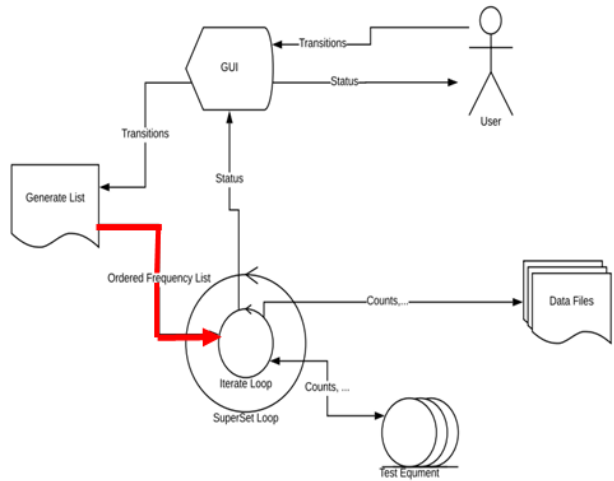

Figure 10 Improved LineShape High Level Diagram via Frequency List Functional Encapsulation

### 3.2.1  Order Flexibility

Concentrating or isolating frequency ordering into one module (Frequency & Power) increases flexibility. User modification of the transition order becomes a useful new feature that provides easy testing of any sequence dependencies that may exist e.g. EOM and detector asymmetries.

This feature is underway as a much needed enhancement that may also be beneficial to the All +1 or -1 modification.

## 3.3 Increased Modularization

In general, modularization allows all the benefits discussed in 3.1 applied to functional scope. Large modules are an indication that subdivision is needed vis a vis functional complexity. LabVIEW IDE "avoids" module bloat by not supporting screen zoom i.e. large networks are difficult to build and maintain. This is another key area where LineShape requires modification – the block diagram occupies 12 standard screens addressed by scrolling vertically and horizontally. Extent reduction can be achieved by increased modularization/sub-vi's or stacking for sequences.

Extent minimization has been implemented in:
- Frequency and Count Unhide
  Much screen real estate was devoted to explicit assignment of control display/visible attributes of center frequency, and count indicators for selected transitions. Contraction was possible by a looping implementation increasing flexibility and further modularized for possible reuse.

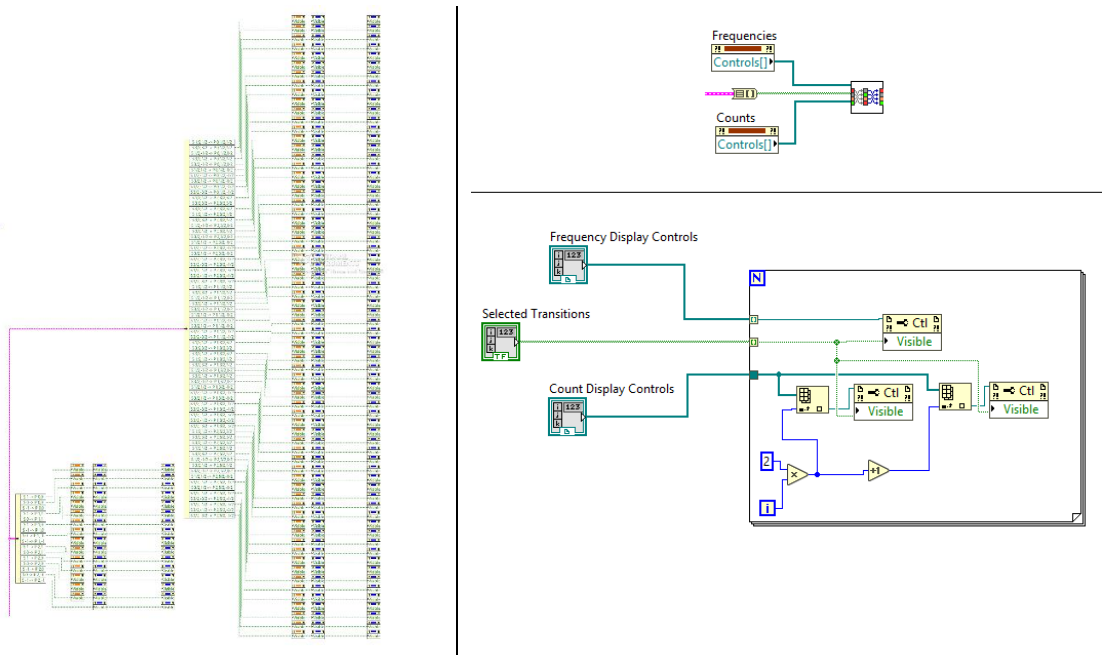

Figure 11 Center Frequency and Count Indicator Hide/Visible Reformation: Original (left), Reformed Segment (right, top), New Module Implementation (right, bottom)

- Inner Loop:
  - Next Frequency
    Extent reduction is possible here by stacked frames or modularization since the functionality was essentially a frame sequence. Modularization was selected to

increase flexibility in timing and equipment options facilitated by increased space for modifications within the new modules.

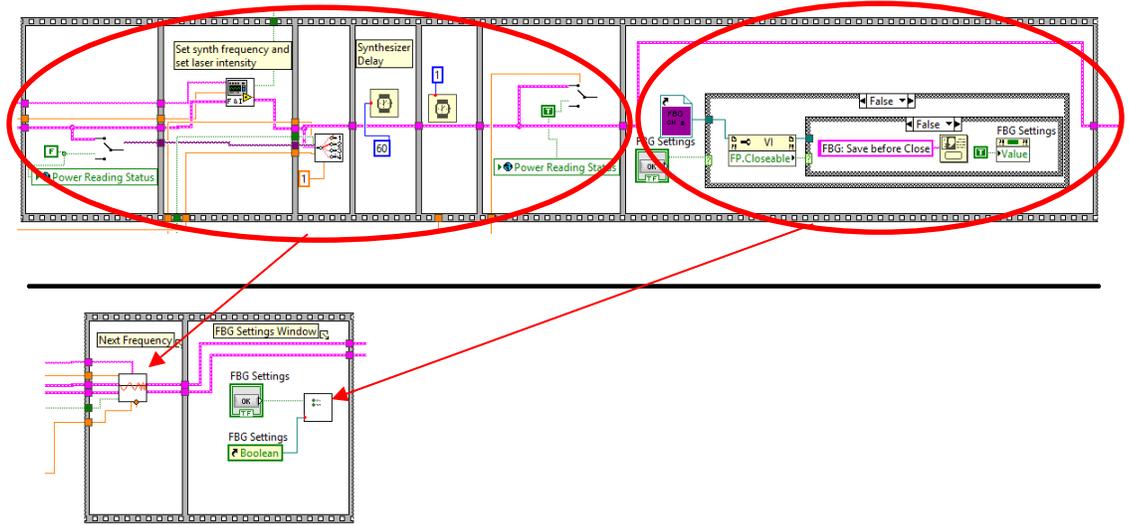

**Figure 12 Next Frequency and FBG Settings Contraction: before (above); after (below)**

- End of Super-set
  Since significant access to local variables occurs in this area sequence stacking was retained while modularizing/sub-vi for long operation chains as in DMM voltage readings. The latter offers both re-use possibility and no use of local variables.

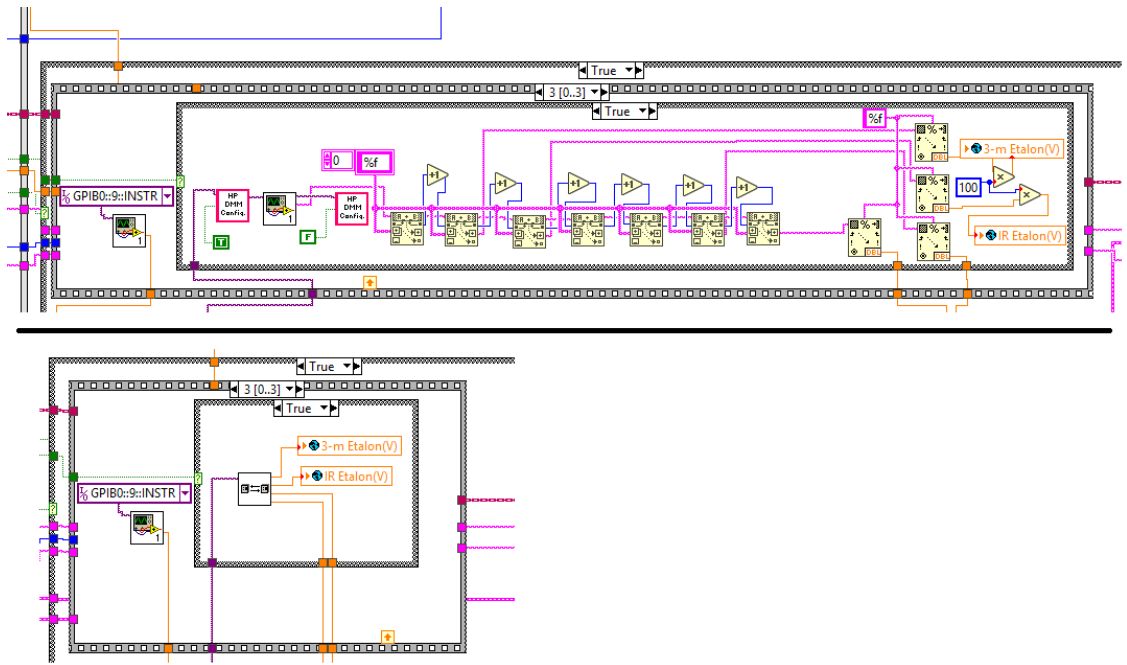

**Figure 13 End of Superset Voltage Gathering: before (above); after (below)**

- Initialization
  As can be seen clearly in Figure 9, further extent reduction can be accomplished in the initialization stage (calls before the superset loop). This has not been implemented to-date.

## 3.4 Polymorphism

Object Oriented Programing (OOP) handles related components/entities with inheritance. A general object can be specialized in a child object inheriting the general methods/functions and data from its parent while adding those specific to its specialization. In search of smaller dead-times, LineShape is experimenting with several different types of counters e.g. GPIB [8] or USB. A Totalizer class was created with inheritance to various specializations as shown in Figure 14.

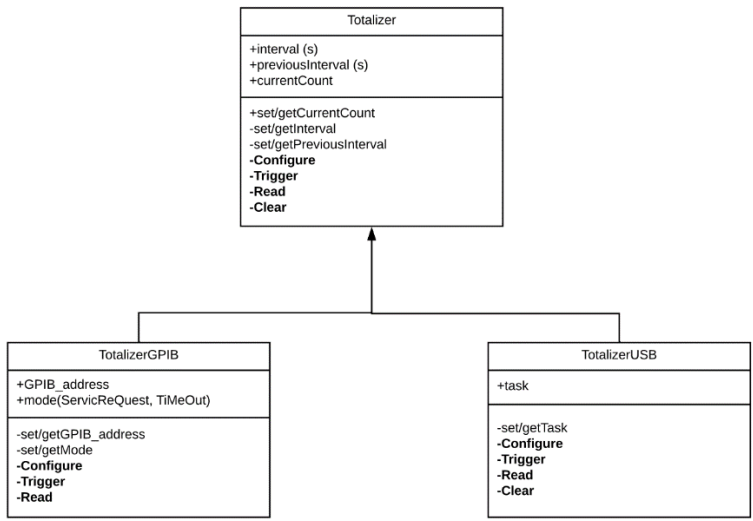

Figure 14 Totalizer Class Hierarchy

Polymorphism is one of the strengths of OOP where the required method/function is selected dynamically based on the specific kind of object making the generic call as shown in Figure 15.

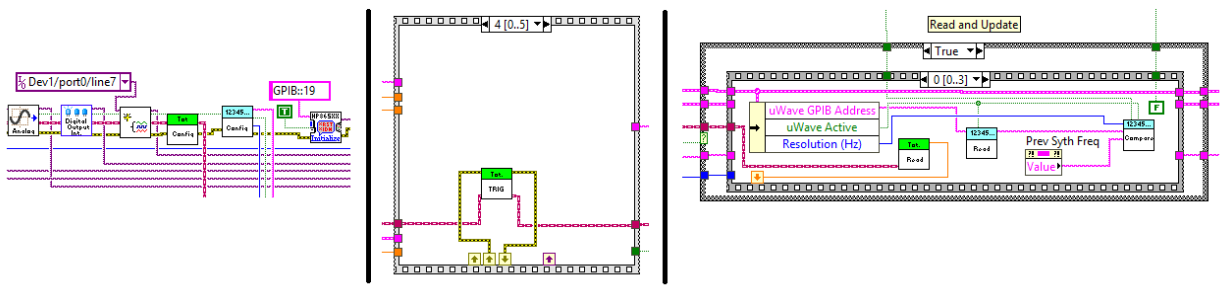

Figure 15 Totalizer Polymorphic Method Calls: Configure (left), Read (middle), Read (right)

This allows flexible programming of generic objects which for example are specialized/selected during configuration. Thus, in our example, adding a new type only involves changes in the configuration module(s) as sketched in Figure 16.

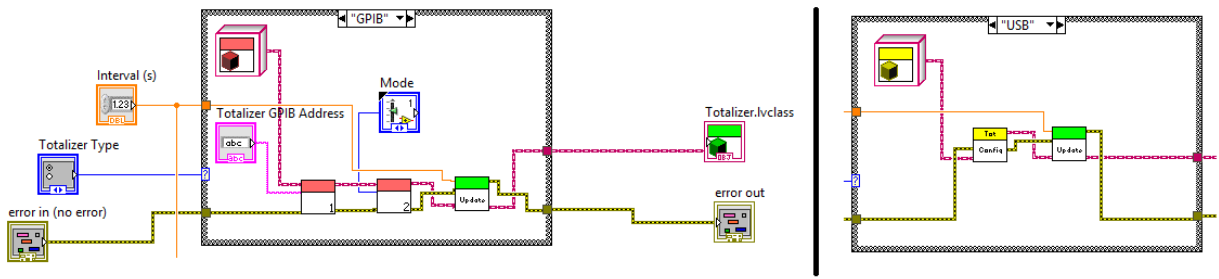

Figure 16 Totalizer Selection: GPIB (left), USB (right)

# 4 Improvement Metrics

Quantifying each attribute allows objective comparison of multi-attribute systems. Weights can be used to assign differential importance to different attributes. We use this approach within each domain.

## 4.1 Environment

As mentioned before, the legacy environment was not fully operation with respect to RAS functionality. Weighted comparison is tabulated in Table 1 for both the broken (Old) and working (W Old) environments to ensure fair contrasts. Survivability is given the heaviest weight. Each environment instance is scored the following observations:

- Backup
  Dynamics duplication offered by a RAS system offers protection against disk failure. But, machine or building catastrophe is not protected.
- Version Control
  While changing names does offer some rollback capability, it introduces fragility in the code base. Every version of sub-component requires a different version of the containing component. Keeping the same name avoids this problem.
- Team Capable
  The legacy environment offered almost no support for multiple team members e.g. conflict protection.

Table 1 Environment Improvement Comparison

| Characteristic | Weight (0-1) | Old Env. (0-1) | Working Old Env. (0-1) | New Env. (0-1) | Advantage (%) | |
|---|---|---|---|---|---|---|
| | | | | | Old | W Old |
| Backup | 0.5 | 0.1 | 0.7 | 1 | 1000 | 143 |
| Version Control | 0.25 | 0.5 | 0.5 | 1 | 200 | 200 |
| Team Capable | 0.25 | 0.1 | 0.1 | 1 | 1000 | 1000 |
| **Overall** | 1.00 | 0.20 | 0.50 | 1.00 | 500 | 200 |

## 4.2 LineShape

It is difficult to quantify objectively the improvement achieved via the strategies deployed without a control study using the original system. Comparison is further exacerbated by possible differing developer programming experience. Only anecdotal testimony is available. However, system performance data is also presented.

Based on current feature addition experience and implementation estimates[7], Table 2 attempts a quantified comparison along the strategies and performance data dimensions. Advantage is the ratio between old and new/improved systems. Specific modifications/features that have been implemented by the author were used as the data points for each strategy as follows:

- Reusable Type: Testing Boolean accessible across all components.
- Functional Encapsulation: Probe alternating sides.
- Increased: Any feature, but modularization generally adds time.
- Polymorphism: Adding another universal counter
- Cycle Time: Total data collection and preparation time i.e. to execute a single frequency.

Table 2 LineShape Improvement Comparison: Feature Addition and Performance

| Modification | Old System (wks) | New System (wks) | Advantage (%) |
| --- | --- | --- | --- |
| Reusable Types | 1 | 0.2 | 500 |
| Functional Encapsulation | 3 | 1 | 300 |
| Increased Modularization | 1 | 1.2 | -17 |
| Polymorphism | 2 | 1 | 200 |
| Cycle Time (ms) | 1000 | 80 | 1250 |
| **Overall Feature Addition** | | | **983** |
| **Overall** | | | **2233** |

Feature addition is therefore seven times faster on average and overall performance is twenty times better.

# 5   Further Work

Looking forward beyond this work several enhancements suggest themselves. While systems continue to evolve as technology does, immediate extension to the environment described can be:

- SVN and Data on disparate GDs
  Dynamic update of Data is preserved while the scheduled GD synchronization is utilized; and
- SVN Server on a separate machine
  Environment durability is increased by distributing functional nodes. SVN Server is a key service that is better hosted on a dedicated machine.

---

[7] Evaluation is based on the author's previous programming experience.

Some of the reformation strategies were not implemented across all instances due to time constraints. Application to all relevant instances for the following strategies is still open:

- Reusable Types Definition completion
  As cited in Section 3.1, File & Equipment, and Input Selection Frequency clusters are ripe for reusable type definition. Only the former has been implemented. Thus the latter and other high usage cluster require attention; and
- Modularization completion
  Further modularization would improve readability of LineShape and reduce the learning curve for new team members. Only initial effort was applied here and much room exists for improvement.